\documentclass[technote,12pt,draftcls,onecolumn,letterpaper]{IEEEtran}

\usepackage{fixltx2e}
\usepackage[cmex10]{amsmath}
\usepackage{amssymb}
\usepackage[font=footnotesize,caption=false]{subfig}
\usepackage{mathtools}
\usepackage{graphicx}
\usepackage{url}
\usepackage{booktabs}
\usepackage[T1]{fontenc}
\usepackage{microtype}
\usepackage{tikz,pgfplots}
\usetikzlibrary{shapes,fit,arrows,decorations.markings,matrix,plotmarks,positioning,fit}
\usepackage{cite}
\usepackage{color}

\pgfplotsset{compat=newest} 

\newtheorem{assump}{Assumption}

\newtheorem{lem}{Lemma}
\newtheorem{prop}{Proposition}
\newtheorem{theorem}{Theorem}
\newtheorem{remark}{Remark}
\newtheorem{defin}{Definition}

\newcommand{\ie}{\emph{i.e.},}

\newcommand{\mc}{\mathcal}
\newcommand{\mbb}{\mathbb}

\renewcommand{\vec}{\mathbf}

\begin{document}

  \title{A One-step Approach to Computing a Polytopic Robust Positively Invariant Set}

  \markboth{IEEE Transactions on Automatic Control}{Trodden: A One-Step Approach to Computing a Polytopic Robust Positively Invariant Set}

  \author{Paul Trodden, \IEEEmembership{Member, IEEE} \thanks{P.~A.~Trodden is with the
  Department of Automatic Control \& Systems Engineering, University
  of Sheffield, Mappin Street, Sheffield S1 3JD, UK (e-mail:
  \texttt{p.trodden@shef.ac.uk}).}}
  
\maketitle
   
  \begin{abstract}
	A procedure and theoretical results are presented for the problem
    of determining a minimal robust positively invariant (RPI) set for
    a linear discrete-time system subject to unknown, bounded
    disturbances. The procedure computes, via the solving of a single
    LP, a polytopic RPI set that is minimal with respect to the family
    of RPI sets generated from a finite number of inequalities with
    pre-defined normal vectors.
  \end{abstract}

\section{Introduction}%
\label{sec:Introduction}%

We consider the problem of finding, for the discrete-time, linear
time-invariant system,
\begin{equation}
x^+ = Ax + w,
\label{eq:dyn}
\end{equation}
a robust positively invariant (RPI) set. That is, a set
$\mc{R} \subset \mbb{R}^n$ with the property 
\begin{equation}
Ax + w \in \mc{R}, \forall x \in \mc{R}, w \in \mbb{W}.
\label{eq:rpi_dyn}
\end{equation}
In this problem, $x \in \mbb{R}^n$ is the current state and $x^+$ its
successor. The disturbance $w \in \mbb{R}^n$ is unknown but lies in a
polytopic (compact and convex) set $\mbb{W}$ that contains the origin
in its interior.

Robust or disturbance invariant sets are important in control, and
their theory and computation have attracted significant attention;
see, for example,~\cite{KG98,Blanchini99,RKK+05,RKM+07} and references
therein. One set that is of particular interest is the \emph{minimal}
RPI (mRPI) set---that is, the RPI set that is smallest in volume among
all the RPI sets for a system---which is also the set of states
reachable from the origin in the presence of a bounded
disturbance. This set is an essential ingredient in many robust
control algorithms. For example, in tube-based robust model predictive
control (MPC)~\cite{MSR05}, an RPI set is used to guarantee robust
stability and feasibility in the presence of bounded uncertainty;
moreover, since the constraints in the MPC optimization problem are
\emph{tightened} according to the size of the RPI set, then the
smallest RPI set (\ie~the mRPI set) is desirable. However, computing
an exact representation of the mRPI is generally impossible (except
for special instances of $A$, as identified later), and instead an
approximation is usually sought. A seminal contribution in this regard
is~\cite{RKK+05}, which proposes a method for computing an abitrarily
close outer-approximation to the mRPI set, which is itself RPI.

The essence of the problem of computing exactly the mRPI set is that
this set is, in general, not finitely determined. Methods for
computing approximations to the mRPI set, including~\cite{RKK+05},
rely on finding finite representations of the set. Recently, in the
context of tube-based MPC, \cite{RKC13} introduced and studied the
notion of a polytopic RPI set defined by a finite number, $r$, of
\emph{a-priori} selected linear inequalities. For a non-autonomous
system $x^+ = Ax + Bu + w$ controlled by a continuous positively
homogeneous control law, $u = \kappa(x)$, the authors showed that the
RPI set dynamic condition~\eqref{eq:rpi_dyn} has an equivalent
representation as $r$ functional inequalities. It was established that
a fixed-point solution to the functional equation corresponds to an
RPI set that is \emph{minimal}, in volume, with respect to the entire
family of RPI sets defined by the pre-selected inequalities, and is an
invariant outer-approximation to the mRPI set. To compute this set,
the authors of \cite{RKC13} give an iterative procedure, based on
solving a sequence of LPs, for which convergence is guaranteed.

In this note, we adopt the notions of~\cite{RKC13} and specialize
their results to the case of the linear autonomous
system~\eqref{eq:dyn} (alternatively, the linear non-autonomous system
with linear state feedback control law) in order to develop a one-step
approach, based on solving a single LP, to the computation of the
smallest RPI set defined by a pre-selected system of
inequalities. Though simple, to the author's knowledge this has not
appeared in the literature, although there are related results; for
example, it is known that checking the invariance of an existing
polytope is an LP~\cite{Blanchini99}. On the other hand, the ability
to synthesize a near-minimal RPI set by solving a single LP
potentially paves the way for robust control methods that re-compute
the disturbance invariant sets on-line, as done in, for example, the
recently developed ``plug-and-play'' approach to distributed
MPC~\cite{RFF14}.

The proposed approach differs to the one of~\cite{RKK+05} in one
important assumption: the number and normal vectors of the
inequalities that represent the RPI set are, as in~\cite{RKC13},
defined \emph{a priori}, while in~\cite{RKK+05} both are unknown until
termination of the algorithm. This \emph{a-priori} definition, first
proposed and studied by~\cite{RKC13}, has two consequences: firstly,
the RPI set obtained is not necessarily the mRPI set, or even an
abitrarily close outer-approximation (as it is in~\cite{RKK+05});
however, it is the smallest RPI set that can be represented by the
finite number, $r$, of chosen inequalities with normal vectors
$\{P_i^\top: i =1\dots r\}$~\cite{RKC13}. To make a clear distinction,
in this note we term this the $(P,r)$-mRPI set when the number of
chosen inequalities is $r$ and the matrix of normal vectors (the
left-hand side of the defining system of inequalities) is
$P$. Secondly, the method of~\cite{RKK+05} involves solving a sequence
of LPs and then computing a Minkowski summation, but here only the
solving of a single LP is required. The development of the procedure
here comprises two steps, the enumeration of which also serves to
clarify the contribution of this note with respect to~\cite{RKC13}:
first, we show that, for the studied linear autonomous
system~\eqref{eq:dyn}, the fixed-point solution to the functional
equation, which~\cite{RKC13} showed is guaranteed to exist, is in fact
unique. Secondly, we show that the corresponding RPI set---which
\cite{RKC13} proved to be minimal with respect to the family of RPI
sets represented by $(P,r)$---can be computed via a single linear
program (LP), as an alternative the iterative sequence of LPs proposed
by~\cite{RKC13}. 

Another method that uses a single LP to compute a disturbance
invariant set is the optimized robust control invariance approach
of~\cite{RKM+07}, applicable to the linear non-autonomous system $x^+
= Ax+Bu+w$. Because a robust \emph{control} invariant (RCI) set---and
the associated control policy---is obtained, then this subsumes the
robust positive invariance (where a \emph{fixed} control law is
assumed) considered here. However, that approach optimizes over only
those control policies that guarantee a finitely determined set,
achieved by employing a relaxed variation of the assumption,
for~\eqref{eq:dyn}, that $A^k\mbb{W} \subseteq \alpha \mbb{W}$ for
some $\alpha \in [0,1)$ and finite integer $k$. In this note, the
  assumption that $A$ has eigenvalues inside the unit circle is
  required, which is different to the assumption used for finite
  determination of RCI sets in~\cite{RKM+07}, but weaker than the
  assumption required for finite determination of the mRPI set
  for~\eqref{eq:dyn}.


The organization of this note is as follows. First, in
Section~\ref{sec:uni}, it is shown that for the system~\eqref{eq:dyn},
the fixed-point solution is, under suitable assumptions,
unique. Subsequently, in Section~\ref{sec:LP}, it is shown that the
$(P,r)$-mRPI set for~\eqref{eq:dyn} may be computed via a single
LP. Finally, examples are given in Section~\ref{sec:ex} to illustrate
the practicality of the proposed approach, before conclusions are made
in Section~\ref{sec:conc}.

\emph{Notation:} The sets of non-negative and positive reals are,
respectively, $\mbb{R}_{0+}$ and $\mbb{R}_+$. For $a, b \in
\mbb{R}^n$, $a \leq b$ applies element by element. A matrix $M$ is
non-negative, denoted $M \geq 0$, if $M_{ij} \geq 0$ for all $i$ and
$j$. $\lambda \mc{X}$ is the scaling of a set $\mc{X}$ by $\lambda \in
\mbb{R}$, defined as $\{\lambda x: x \in \mc{X}\}$.  $A\mc{X}$ denotes
the image of a set $\mc{X} \subset \mbb{R}^n$ under the linear map $A
: \mbb{R}^n \mapsto \mbb{R}^p$, and is given by $\{ Ax : x \in
\mc{X}\}$. The support function of a set $\mc{X}$ is $h(\mc{X},v)
\triangleq \sup \{ v^\top x: x \in \mc{X} \}$.  A polyhedron is the
convex intersection of a finite number of halfspaces, and a polytope
is a closed and bounded (hence compact) polyhedron.

\section{Existence and uniqueness of a $(P,r)$-mRPI set}%
\label{sec:uni}
For the system~\eqref{eq:dyn}, we consider the case of a polytopic 
disturbance set
\begin{equation}
\mathbb{W} \triangleq \bigl\{ w \in \mbb{R}^n : Fw \leq g \bigr\},
\label{eq:W}
\end{equation}
where $F \in \mbb{R}^{p \times n}$, $g \in \mbb{R}^p_{0+}$, and make
the following two standing assumptions.
\begin{assump}\label{assump:W}
The set $\mbb{W}$ contains the origin in its interior.
\end{assump}
\begin{assump}\label{assump:A}
The eigenvalues of $A$ are strictly within the unit circle.
\end{assump}

The former assumption requires that $g \in \mbb{R}^p_{+}$. The latter
assumption implies, as shown in~\cite{KG98}, that for a given compact
disturbance set $\mbb{W}$ there exists a compact RPI set, $\mc{R}$,
for the system~\eqref{eq:dyn}, satisfying~\eqref{eq:rpi_dyn}.
\begin{assump}\label{assump:R}
The RPI set $\mc{R}$ is a polytope that contains the origin in
its interior.
\end{assump}

Note that Assumptions~\ref{assump:W} and~\ref{assump:R} imply that the
support functions to $\mbb{W}$ and $\mc{R}$, respectively, are
positive---a key technical property that will be used in this note to
establish the existence and uniqueness of the RPI set that we aim to
compute.

In this note, following~\cite{RKC13}, we consider the RPI set
constructed from a finite number, $r$, of inequalities with
pre-defined normal vectors. That is, $\mc{R}\triangleq \mc{R}(q)$,
defined as
\begin{equation}
\mathcal{R}(q) \triangleq \bigl\{ z \in \mbb{R}^n : Pz \leq q \bigr\}, 
\label{eq:R}
\end{equation}
where $P \in \mbb{R}^{r \times n}$, $\bigl\{ P_i^\top : i \in
\{1,\dots,r\}\bigr\}$ spans $\mbb{R}^n$, $P_i$ is the $i$th row of
matrix $P$, and $q \in \mbb{R}^r_{0+}$. The left-hand side of the
inequalities---the matrix $P$---is to be chosen \emph{a priori} by the
designer. The following result, which is an application of Farkas'
Lemma, establishes basic conditions on the matrices $A$, $P$ and $F$
for the existence of an RPI set for the system~\eqref{eq:dyn} given
the disturbance polytope~\eqref{eq:W}.
\begin{theorem}[Adapted from Hennet and Castellan~\cite{HC01}]\label{thm:Hennet}
Suppose Assumptions~\ref{assump:W}--\ref{assump:R} hold. Then the set
$\mc{R}(q)$ with some $q=\bar{q}$ is robust positive invariant for the
system~\eqref{eq:dyn} if and only if there exist non-negative matrices
$H \in \mbb{R}^{r\times r}$ and $M \in \mbb{R}^{r \times p}$ such that
\begin{subequations}
\begin{align}
HP &= PA\\
MF &= P\\
H \bar{q} + M g &\leq \bar{q}
\end{align}
\end{subequations}
\end{theorem}

We will assume that $P$ is chosen so that an RPI set exists:
\begin{assump}\label{assump:P}
For the chosen $P$, and the system $(A,\mathbb{W})$, there exists
a $\bar{q} \in \mbb{R}^r_{+}$ such that \eqref{eq:rpi_dyn} holds for
all $x \in \mc{R}(\bar{q})$.
\end{assump}
\begin{remark}
While Assumption~\ref{assump:P} may appear strong, it is needed
to narrow the class of matrices that we consider to those that admit
an RPI set. However, the procedure presented in the next section
includes a easy certification of existence of an RPI set for a chosen
$P$: if an RPI set exists, the $(P,r)$-mRPI set is returned. If no RPI set
exists, the optimization problem is unbounded.
\end{remark}

The authors of \cite{RKC13} show---in the more general setting of a linear
non-autonomous system controlled by a positively homogeneous
state-feedback control law---that RPI condition~\eqref{eq:rpi_dyn} is
equivalent to the functional inequality~
\begin{equation}
c(q) + d \leq b(q),
\label{eq:fp_eq}
\end{equation}
where, for $i = 1\dots r$, $b_i(q) \triangleq h(\mc{R}(q),P_i^\top)$,
$c_i(q) \triangleq h(A\mc{R}(q),P_i^\top)$, $d_i \triangleq
h(\mbb{W},P_i^\top)$. That is, the set inclusion requirement is
replaced by support function inequalities, which is a standard
technique~\cite{Schneider93}. Note that $b(q)$ may be different to
$q$; for example, in the case of redundant inequalities defining
$\mc{R}(q)$. The topological properties of these functions described
in the following two lemmas are essential to establishing existence
and uniqueness of the fixed-point solution to~\eqref{eq:fp_eq}.
\begin{lem}[Adapted from Proposition 1 of~\cite{RKC13}]\label{lem:bcd} 
Suppose that Assumptions~\ref{assump:W}--\ref{assump:R} hold. Then the
functions $b \colon \mbb{R}^r_{0+} \mapsto \mbb{R}^r_{0+}$, $c \colon
\mbb{R}^r_{0+} \mapsto \mbb{R}^r_{0+}$ are continuous and
monotonically non-decreasing; that is, $b(a_1) \leq b(a_2)$ for $a_1
\leq a_2$. Also, $d \in \mbb{R}^r_{+}$.
\end{lem}

\begin{lem}[Positive homegeneity of $b$, $c$]
Suppose Assumptions~\ref{assump:A} and~\ref{assump:R} hold. Then the
functions $b(\cdot)$ and $c(\cdot)$ are positively homogeneous; that
is $b(\lambda a) = \lambda b(a)$ for $\lambda \geq 0$, with a similar
expression for $c(\cdot)$.
\end{lem}
\begin{IEEEproof}
Consider $b_i(\lambda a) = h\left(\mc{R}(\lambda a),P_i\right)$ for
some $a \in \mbb{R}^r_{0+}$, $\lambda \geq 0$ and $i \in
\{1,\dots,r\}$. By definition of $\mc{R}(\cdot)$, $\mc{R}(\lambda a) =
\lambda \mc{R}(a)$. Thus, $h\left(\mc{R}(\lambda a),P_i\right) =
h\left(\lambda\mc{R}(a),P_i\right) = \lambda
h\left(\mc{R}(a),P_i\right)$, for $\lambda \geq 0$, where the latter
equality follows directly from the definition of the support
function~\cite{Schneider93}. Hence, $b_i(\lambda a) = \lambda b_i(a)$,
therefore $b(\lambda a) = \lambda b(a)$. Positive homogeneity of
$c(\cdot)$ may be established by the same arguments.
\end{IEEEproof}

The next result, which concerns the existence of a fixed-point
solution to~\eqref{eq:fp_eq}, was established by~\cite{RKC13} in the
setting of a linear non-autonomous system controlled by positively
homogeneous state-feedback control law, and hence immediately applies
to the more specialized case considered in this note.

\begin{theorem}[Theorem 1 of \cite{RKC13}]\label{thm:RKC}
Suppose Assumptions~\ref{assump:W}--\ref{assump:R} hold. Let $\mc{Q}
\triangleq \bigl\{ q \in \mbb{R}^r_{0+} : 0 \leq q \leq \bar{q}
\bigr\}$. Then, (i) for all $q \in \mc{Q}$, $c(q) + d \in \mc{Q}$ and
(ii) there exists at least one $q^* \in \mc{Q}$ satisfying $c(q^*) + d
= b(q^*) = q^*$ if and only if Assumption~\ref{assump:P} holds.
\end{theorem}
\begin{remark}
The necessity and sufficiency of Assumption~\ref{assump:P} follows by
definition. In particular, if Assumption~\ref{assump:P} does not hold,
then there does not exist an RPI set for the system $(A,\mathbb{W})$
with the chosen $P$.
\end{remark}

\begin{remark}\label{rem:pos}
Note that, in view of the assumptions on $g$ and the properties of
$b(\cdot)$, $c(\cdot)$, and $d$, a fixed-point solution $q^*$ must be
strictly positive.
\end{remark}

With respect to computing a fixed-point solution, the sequence
generated by the iterative procedure $q^{[p+1]} = c(q^{[p]}) + d$,
with $q^{[0]} = 0$, converges to the fixed-point solution $q^*$ with
the smallest 1-norm value, $\|q^*\|_1$~\cite[Theorem 2]{RKC13}. As the
following result states, the corresponding set $\mc{R}(q^*)$ is RPI,
and, in fact, is the minimal (smallest volume) RPI set over the family
of RPI sets defined by the $r$ inequalities with left-hand side $P$.
\begin{lem}[Corollary 1 of \cite{RKC13}]
$\mc{R}(q^*) = \bigcap_{X \in \mc{S}} X$ where
\begin{equation*}
\mc{S} \triangleq \left\{ \mc{R}(q) : q \in \mc{H} \right\}, \ \text{and} \ \mc{H} \triangleq \left\{ q \in \mbb{R}^r_{0+} : c(q) + d \leq b(q) \right\}
\end{equation*}
\end{lem}

For convenience, we define this set $\mc{R}(q^*)$ as the $(P,r)$-mRPI set.

\begin{defin}[$(P,r)$-mRPI set]
The $(P,r)$-mRPI set for system~\eqref{eq:dyn} is $\mc{R}(q^*)$ where $q^* = b(q^*) = c(q^*)+d$.
\end{defin}

In this note, we propose an alernative to the iterative procedure
of~\cite{RKC13}. To this end, the next result shows that the
fixed-point solution to~\eqref{eq:fp_eq} is, in fact, unique. This
result is then exploited in Section~\ref{sec:LP}, wherein the problem
of finding the fixed-point solution is cast as an LP.

\begin{theorem}[Uniqueness of fixed-point solution]\label{thm:unique}
Suppose Assumptions \ref{assump:W}--\ref{assump:P} hold. Then there
exists a unique $q^* \in \mbb{R}^r_{+}$ satisfying $c(q^*) + d =
b(q^*) = q^*$.
\end{theorem}

\begin{IEEEproof}
Existence is established by Theorem~\ref{thm:RKC}, so it remains to
show that $q^*$ is unique. Let $l(q) = c(q) + d - b(q)$ and $f(q) =
b(q) - q$. Finding the fixed-point solution $c(q^*) + d = b(q^*) =
q^*$ is equivalent to finding $q^*$ such that $l(q^*) = f(q^*) =
0$. Suppose there exist $q^1 \in \mbb{R}^r_+$ and $q^2 \in
\mbb{R}^r_+$ such that $l(q^1) = f(q^1) = 0$, $l(q^2) = f(q^2) =
0$, and $q^2 \neq q^1$, \ie~$q^2 - q^1 \neq 0$. There are two possibilities:
\begin{enumerate}
\renewcommand{\theenumi}{(\roman{enumi})}
\renewcommand{\labelenumi}{\theenumi}
\item\label{case:a} $q^2_i > q^1_i$ for at least one $i \in \{1,\dots,r\}$, with
  $q^2_j \leq q^1_j$ otherwise;
\item\label{case:b} $q^2 \leq q^1$, with $q^2_i < q^1_i$ for at least
  one $i \in \{1,\dots,r\}$.
\end{enumerate}
Consider case~\ref{case:a}. Let
\begin{equation*}
\alpha = \min_{i=1\dots r} \left\{ \frac{q^1_i}{q^2_i} \right\}= \frac{q^1_p}{q^2_p} > 0
\end{equation*}
Strict positivity follows from the discussion in
Remark~\ref{rem:pos}. Since $q^2_i > q^1_i$ for at least one $i$, then
$\alpha < 1$. Let $s = \alpha q^2 < q^2$. It follows, from positive
homogeneity of $b(\cdot)$ and the fact that $b(q^2)-q^2 =0$, that
$f(s) = b(s) - s = b(\alpha q^2) - \alpha q^2 = \alpha \bigl(b(q^2) -
q^2)\bigr) = 0$. Similarly,
\begin{equation*}
\begin{split}
l(s) &= c(s) + d - b(s) \\
&= c(\alpha q^2) + d - b (\alpha q^2)\\
&= \alpha c(q^2) + d - \alpha b (q^2)\\
&= \alpha \bigl( c(q^2) - b(q^2) \bigr) + d \\
&> 0
\end{split}
\end{equation*}
where the second line follows from the positive homegeneity of
$c(\cdot)$ and $b(\cdot)$, and the strict inquality with zero follows
from $c(q^2)-b(q^2)=-d$, $\alpha < 1$ and $d > 0$. Now, by definition
of $\alpha$, and since $\alpha < 1$, then $s \leq q^1$ with $s_p =
q^1_p$. For the same $p$, we have $f_p(q^1) = b_p(q^1) - q^1_p = 0$,
$f_p(s) = b_p(s) - s_p=0$, and, since $s \leq q^1$, then $b_p(s) \leq
b_p(q^1)$. In fact, $b_p(s) = b_p(q^1)$, as we have already shown that
$b_p(s) = s_p = q^1_p$. We also have $l_p (q^1) = c_p(q^1) + d_p -
b_p(q^1)$ and $l_p (s) = c_p(s) + d_p - b_p(s)$. Because $b_p(q^1) =
b_p(s)$ and $c_p(s) \leq c_p(q^1)$, it follows that $l_p(s) \leq
l_p(q^1)$. But then $0 = l_p(q^1) \geq l_p(s) > 0$, and we have a
contradiction: therefore, we conclude that case~\ref{case:a} cannot
hold, and either case~\ref{case:b} holds or $q^2 = q^1$. Now consider
case~\ref{case:b}, and its equivalent statement: $q^1_i > q^2_i$ for
at least one $i \in \{1,\dots,r\}$, with $q^1_j \geq q^2_j$
otherwise. Following the same set of arguments, starting with the
opposite definitions of $\alpha =\min_{i=1\dots r} \left\{ q^2_i/q^1_i
\right\} $ and $s=\alpha q^1$, we find that that case~\ref{case:b}
cannot hold either. Therefore, $q^1 = q^2 = q^*$, and the solution is
unique.
\end{IEEEproof}

\section{Computing the $(P,r)$-mRPI set via a single LP}
\label{sec:LP}

The problem of computing the $(P,r)$-mRPI set is that of finding the
$q$ that satisfies the functional inequality (RPI condition) (6) while
attaining the smallest value of $\|q\|_1$. The results in the previous
section show that this $q$ in fact satisfies (6) with equality; it is
the fixed-point solution $q^*$. Therefore, the problem of finding
$q^*$ is
\begin{equation}
q^* = \arg \min_{q} \left\{ \left\| q \right\|_1 : c(q) + d \leq b(q) \right\}
\label{eq:minimax}
\end{equation}
This is not tractable, as, by the definitions of $b(\cdot)$ and
$c(\cdot)$, the constraints are maximization problems involving the
optimization variable:
\begin{multline*}
\max \left\{ P_i A x : Px \leq q \right\} + \max \left\{ P_i w : Fw \leq g \right\} \\\leq \max \left\{ P_i x : Px \leq q \right\}
\end{multline*}
for $i=1\dots r$. However, by noting that the fixed-point solution is
unique, we may replace the problem of~\eqref{eq:minimax} with the
maximization problem
\begin{equation*}
q^* = \arg\max_{q} \left\{ \left\| q \right\|_1 : c(q) + d = b(q) \right\}
\end{equation*}
This problem then easily converts to a linear program, as shown by
the following. Introduce auxiliary variables $\xi^i \in \mbb{R}^n$ and
$\omega^i \in \mbb{R}^n$ for each RPI inequality $i \in
\{1,\dots,r\}$. Then, noting that $q = b(q) = c(q) + d$ at the desired
fixed-point solution, eliminate $q$ and $b(q)$, leading to the problem
\begin{equation}
\mathbb{P}\colon q^* = c^* + d^*, \ \text{where} \ (c^*,d^*) =
\arg\max_{\substack{\{c_i, d_i, \xi^i, \omega^i\}\\\forall i \in
    \{1,\dots,r\}}} \sum_{i=1}^r c_i + d_i
\label{eq:Pcost}
\end{equation}
subject to, for all $i \in \{1,\dots,r\}$,%
\begin{subequations}
\begin{align}
c_i &\leq P_i A \xi^{i},\label{eq:cubound}\\
P \xi^i &\leq c + d,\label{eq:clbound}\\
d_i &\leq P_i \omega^{i},\label{eq:dbound}\\
F \omega^i &\leq g.\label{eq:wbound}
\end{align}
\label{eq:primalcons}
\end{subequations}
In this problem, maximizing each $c_i$ subject to
constraints~\eqref{eq:cubound} and \eqref{eq:clbound} represents finding
the vector of support functions to
$A\mc{R}$. Constraint~\eqref{eq:clbound} represents $Px \leq b(q)$,
with the condition $c(q)+d = b(q)$
enforced. Constraints~\eqref{eq:dbound} and \eqref{eq:wbound} represent
finding $d$, the vector of support functions to $\mbb{W}$.

\begin{remark}
Note that, by definition, $d_i = h(\mbb{W},P_i^{\top})$ is constant and does
not depend on $q$. Therefore, $d$ could be computed prior to solving
$\mbb{P}$, by solving a sequence of LPs, before entering the
optimization as a parameter. However, our aim is to formulate a single
LP (a one-step procedure) that computes, simultaneously, $d$, $c$ and hence $q$.
\end{remark}

Note that each $d_i$ and $\omega_i$ is bounded, via~\eqref{eq:dbound}
and~\eqref{eq:wbound} and the assumptions on $\mathbb{W}$. Further
note that this problem always has a \emph{feasible} solution, since
one can choose, for example, $c_i = d_i = 0$ and $\xi^i = \omega^i =
0$. The question, then, is whether an optimal solution exists, or the
problem is unbounded. To this end, we require the following result,
which specializes Theorem~\ref{thm:Hennet} to the fixed-point solution.
\begin{prop}\label{prop:fp}
%
Suppose Assumptions~\ref{assump:W}--\ref{assump:P} hold. A vector
$q^*$ satisfies the fixed-point relation $c(q^*) + d = b(q^*)= q^*$ if
and only if there exist non-negative matrices $H \in \mbb{R}^{r\times
  r}$ and $M \in \mbb{R}^{r \times p}$ such that
\begin{subequations}
\begin{align}
HP &= PA\label{eq:HP}\\
MF &= P\label{eq:MF}\\
H q^* + M g &= q^*
\end{align}
\label{eq:newHMq}
\end{subequations}
\end{prop}
\begin{IEEEproof}
Consider the $i$th element of each of $c(q^*)$, $d$ and $b(q^*)$,
defined by the (primal) LPs
\begin{subequations}
\begin{align}
c_i(q^*) &= \max \bigl\{ P_i A x: Px \leq q^*\bigr\}\\
d_i &= \max \bigl\{ P_iw : Fw \leq g \bigr\}\\
b_i(q^*) &= \max \bigl\{ P_i x: Px \leq q^*\bigr\}
\end{align}
\label{eq:terms}%
\end{subequations}
If Assumptions~\ref{assump:W}--\ref{assump:P} hold, then by the previous results there exists a $q^*$ satisfying the fixed-point equation. Moreover, each of the terms in~\eqref{eq:terms} is well defined, which is the case if and only if each LP is feasible and attains an finite optimum. Therefore, by weak duality, the dual of each LP
\begin{align*}
c_i(q^*) &\colon \min \bigl\{ h_i^\top q^*: h_i^\top P = P_i A, h_i \geq 0\bigr\},\\
d_i &\colon \min \bigl\{ m_i^\top g : m_i^\top F = P_i, m_i \geq 0\bigr\},\\
b_i(q^*) &\colon \min \bigl\{ y_i^\top q^*: y_i^\top P = P_i, y_i \geq 0\bigr\},
\end{align*}
is feasible. Examining these dual problems, dual feasible solutions
exist if and only if there exist non-negative $h_i \in \mbb{R}^r$,
$m_i \in \mbb{R}^p$, $y_i \in \mbb{R}^r$ such that
\begin{align*}
h_i^\top P &= P_i A, \\
m_i^\top F &= P_i,\\
y_i^\top P &= P_i.
\end{align*}
Applying strong duality, which holds in view of the previous
arguments, to each of the three LPs
\begin{align*}
c_i(q^*) &= h_i^\top q^*,\\
d_i &= m_i^\top g,\\
b_i(q^*) &= y_i^\top q^*. 
\end{align*}
Collecting all rows $i=1\dots r$,
\begin{align*}
c(q^*) &= H q^*\\
d &= M g,\\
b(q^*) &= Y q^*, 
\end{align*}
where $HP = PA$, $MF = P$, $YP = P$. Therefore, it follows that if the
fixed-point equation
\begin{align*}
c(q^*) + d = b(q^*) = q^*
\end{align*}
is satisfied, then so are the conditions~\eqref{eq:newHMq};
conversely, if \eqref{eq:newHMq} are satisfied, then so is the
fixed-point equation.
\end{IEEEproof}

Then the main result of this section follows.

\begin{theorem}
Suppose Assumptions~\ref{assump:W}--\ref{assump:R} hold. If $P$
satisfies Assumption~\ref{assump:P}, then problem $\mathbb{P}$ admits
an optimal solution corresponding to the fixed-point solution
$q^*$. Otherwise, $\mathbb{P}$ is unbounded above.
\end{theorem}

\begin{IEEEproof}
We use duality to prove the theorem. Our goal is to prove that the
optimal solution to $\mbb{P}$ satisfies the
conditions~\eqref{eq:newHMq}, for some non-negative $H$ and $M$, if
and only if Assumption~\ref{assump:P} holds, and that $\mbb{P}$ is
otherwise unbounded. Since the primal LP problem $\mathbb{P}$ is
known to be feasible, it suffices to show that the dual problem is
feasible---and the solution is as claimed---if and only if $P$
satisfies Assumption~\ref{assump:P}; on the other hand, if the dual is
infeasible, then by weak duality the primal problem $\mathbb{P}$ is
unbounded.

The dual problem is
\begin{equation}
\mathbb{D}\colon \min_{\substack{\{\lambda_i, \nu_i, \mu^i, \eta^i\}\\\forall i \in \{1,\dots,r\}}} \sum_{k=1}^r (\eta^{k})^\top g
\end{equation}
subject to, for all $i \in \{1,\dots,r\}$,%
\begin{subequations}
\begin{align}
\lambda - \sum_{k=1}^r \mu^k &= \mathbf{1},\label{eq:dual1}\\
\nu - \sum_{k=1}^r \mu^k &= \mathbf{1},\label{eq:dual2}\\
P^\top \mu^i - A^\top P_i^\top \lambda_i &= 0,\label{eq:dual3}\\
F^\top \eta^i - P_i^\top \nu_i &= 0,\label{eq:dual4}\\
\lambda_i, \nu_i &\geq 0\\
\mu^i, \eta^i &\geq 0
\end{align}
\label{eq:dualcons}%
\end{subequations}
where $\lambda_i \in \mbb{R}$, $\mu^i \in \mbb{R}^r$, $\nu_i \in
\mbb{R}$, $\eta^i \in \mbb{R}^p$ are the dual variables associated
with constraints~\eqref{eq:cubound}--\eqref{eq:wbound}
respectively. 

We first suppose the dual problem $\mbb{D}$ is feasible.
From~\eqref{eq:dual1} and~\eqref{eq:dual2}, $\lambda_i = \nu_i = 1 +
\sum_{k=1}^{r} \mu^k_i$, for all $i=1,\ldots,r$, where $\mu^k_i$ is
the $i$th element of $\mu^k \in \mbb{R}^r$. From this
and~\eqref{eq:dual3}, \eqref{eq:dual4}, it follows that
\begin{align*}
P_i A = \frac{(\mu^i)^\top}{1+\sum_{k=1}^{r} \mu^k_i} P,\\
P_i = \frac{(\eta^i)^\top}{1+\sum_{k=1}^{r} \mu^k_i} F,
\end{align*}
where the division is permitted since $\sum_{k=1}^r \mu^k_i \geq
0$. Collecting all rows $i=1\dots r$ of $P$, it follows that a
\emph{feasible} solution to $\mbb{D}$ satisfies~\eqref{eq:HP}
and~\eqref{eq:MF} with $H_{ij} = \mu^i_j / (1+\sum_{k=1}^{r} \mu^k_i)
\geq 0$, $j=1\dots r$, $M_{ij} = \eta^i_j /( 1+\sum_{k=1}^{r} \mu^k_i)
\geq 0$, $j=1\dots p$; therefore, $H$ and $M$ are non-negative
matrices.


Now we study the \emph{optimal} solution to $\mbb{D}$. Since the
primal problem $\mbb{P}$ is known to be feasible, and we assumed
$\mbb{D}$ to be feasible, then by strong duality (which holds
regardless of the feasibility of $\mbb{D}$) the optimal solutions to
$\mbb{P}$ and $\mbb{D}$ are attained and equal in objective value. So,
applying complementary slackness to \eqref{eq:cubound}
and~\eqref{eq:dbound},
\begin{align*}
\sum_{i=1}^r \lambda^*_i (c_i^* - P_i A \xi^{i*}) &= 0\\
\sum_{i=1}^r \nu^*_i (d_i^* - P_i \omega^{i*}) &= 0.
\end{align*}
where ${}^*$ denotes a variable in the optimal solution. Since each
term in these sums is non-positive,
\begin{align*}
\lambda^*_i (c_i^* - P_i A \xi^{i*}) &= 0,\\
\nu^*_i (d_i^* - P_i \omega^{i*}) &= 0,
\end{align*}
for $i=1,\dots,r$. Morever, because (by \eqref{eq:dual1}
and~\eqref{eq:dual2}) $\lambda^*_i > 0$ and $\nu^*_i > 0$, then, at the optimum,
\begin{align*}
c_i^* &= P_i A \xi^{i*},\\
d_i^* &= P_i \omega^{i*}.
\end{align*}
Hence,
\begin{equation}
\begin{split}
c_i^* + d_i^* &= P_i A \xi^{i*} + P_i \omega^{i*} \\ &= \frac{(\mu^{i*})^\top}{1+\sum_{k=1}^{r} \mu^{k*}_i} P \xi^{i*} + \frac{(\eta^{i*})^\top}{1+\sum_{k=1}^{r} \mu^{k*}_i} F \omega^{i*} \\
& = \frac{1}{1+\sum_{k=1}^{r} \mu^{k*}_i} \left( \sum_{j=1}^r \mu_j^{i*} P_j \xi^{i*} + \sum_{j=1}^p \eta_j^{i*} F_j \omega^{i*} \right).
\end{split}
\label{eq:cplusd}
\end{equation}
Now consider the inequality~\eqref{eq:wbound}. Suppose $F \omega^{i*}
< g$ for some $i \in \{1,\ldots,r\}$ (\ie~$F_j \omega^{i*} < g_j$ for
all $j=1\ldots p$). Complementary slackness implies that $\eta^{i*} =
0$ which in turn implies (from~\eqref{eq:dual4}, assuming that $P_i$
is not trivially all zeros) that $\nu^*_i = 0$; but $\nu^*_i \geq
1$ by~\eqref{eq:dual2}, which is a contradiction. Hence, there must
exist a subset $\mc{K} \subset \{1,\ldots,p\}$ of active constraints
for which $F_k \omega^{i*} = g_k$ for $k \in \mc{K}$. But for any $j
\notin \mc{K}$, $\eta^{i*}_j=0$.

Similarly, consider the inequality~\eqref{eq:clbound}.  By
complementary slackness, if $P \xi^{i*} < c^* + d^*$ then
$\mu^{i*}=0$. By~\eqref{eq:dual3}, this implies that $A^\top P_i^\top
\lambda^*_i = 0$.  There are two cases to consider: (i) if any
elements of $P_i A$ are non-zero then $\lambda^*_i = 0$; (ii) if $P_i
A = 0$ then $\lambda^*_i > 0$ is permitted.  We leave case (ii) for
now and consider (i) first.  $\lambda^*_i = 0$
contradicts~\eqref{eq:dual1}, which requires $\lambda_i^* \geq
1$. Hence, there must exist a subset $\mc{J} \subset \{1,\ldots,r\}$
of active constraints for which $P_j \xi^{i*} = c_j^* + d_j^*$ for $j
\in \mc{J}$.  But for any $k \notin \mc{J}$, $\mu^{i*}_k=0$. As a
consequence of the preceding arguments, \eqref{eq:cplusd} may be
re-written as
\begin{equation*}
\begin{split}
c_i^* + d_i^* &= \frac{1}{1+\sum_{k=1}^r \mu^{k*}_i} \left( \sum_{j \in \mc{J}} \mu_j^{i*} P_j \xi^{i*} + \sum_{k \in \mc{K}} \eta_k^{i*} F_k \omega^{i*}\right)\\
&= H_i (c^* + d^*) + M_i g
\end{split}
\end{equation*}
where $H_i$ is the $i$th row of $H$ and $M_i$ is the $i$th row of
$M$. The second line follows because $H_{ij} = 0$ for $j \notin
\mc{J}$ and $M_{ik} = 0$ for $k \notin \mc{K}$, while $P_j \xi^{i*} =
c_j^* + d_j^*$ for $j \in \mc{J}$ and $F_k \omega^{i*} = g_k$ for $k
\in \mc{K}$.

Now case (ii). If $P_i A = 0$ then $c_i^* = 0$. Moreover, $\lambda_i^*
\geq 1$ is permitted, so the same contradiction is not
constructed. Then, however, either $P \xi^{i*} < d^*$, hence $\mu^{i*}
= 0$, or $P \xi_j^{*i} = d_j^*$, with $\mu^{i*}_j \geq 0$, for $j \in
\mc{J} \subset \{1,\dots,r\}$, and $\mu^{i*}_k = 0$ for all $k \notin
\mc{J}$. Either way,
\begin{equation*}
c_i^* + d_i^* = H_i(c^* + d^*) + M_i g
\end{equation*}
as before. 

Finally, collecting all rows $i=1\ldots r$,
\begin{equation*}
H(c^* + d^*) + Mg = c^* + d^*
\end{equation*}
which is the third condition in~\eqref{eq:newHMq}. This establishes
that the solution to $\mbb{P}$, if it is attainable, satisfies the
conditions~\eqref{eq:newHMq} for it to be the fixed-point solution. It
is attainable if and only if the dual problem $\mbb{D}$ is
feasible. Therefore, it remains to show that the $\mbb{D}$ is feasible
if and only if Assumption~\ref{assump:P} holds.

First, necessity of Assumption~\ref{assump:P}. Suppose
Assumption~\ref{assump:P} is not satisfied, but the dual $\mbb{D}$ is
feasible. By definition, if Assumption~\ref{assump:P} is not satisfied
then for the chosen $P$ and system $(A,\mbb{W})$ there does not exist
a $q$ satisfying the functional
inequality~\eqref{eq:fp_eq}. Therefore, there exists no $q^*$
satisfying the functional equation and, by Proposition~\ref{prop:fp},
the conditions~\eqref{eq:newHMq}. However, the attainable optimal
solution to $\mbb{P}$ and $\mbb{D}$ satisfies~\eqref{eq:newHMq} with
non-negative $H$ and $M$, as has been shown. Therefore, we have a
contradiction, and conclude the optimal solution is attainable, and
$\mbb{D}$ is feasible, only if Assumption~\ref{assump:P} holds.

Second, sufficiency of Assumption~\ref{assump:P}. Writing the primal
constraints~\eqref{eq:primalcons} in the form $\vec{A}\vec{x} \leq
\vec{b}$, where $\vec{x}$ is the vector of primal decision variables,
it follows that the dual constraints~\eqref{eq:dualcons} may be
written in the form $\vec{A}^\top \vec{y} = \vec{c}$, $\vec{y} \geq
0$, where $\vec{y}$ is the vector of dual variables and $\vec{c}$ is
the coefficients vector in the vectorized form, $\vec{c}^\top
\vec{x}$, of the objective function~\eqref{eq:Pcost}. By Farkas'
Lemma, a feasible solution to $\vec{A}^\top \vec{y} = \vec{c}$,
$\vec{y} \geq 0$ exists if and only if $\vec{A}\vec{x} \geq 0 \implies
\vec{c}^\top \vec{x}\geq 0$.  Hence, we aim to show that, if
Assumption~\ref{assump:P} holds, then for all $\vec{x}$ satisfying
$\vec{A}\vec{x} \geq 0$ we also have $\vec{c}^\top \vec{x} \geq
0$. The system $\vec{A}\vec{x} \geq 0$ may be written in terms of the
primal variables as
\begin{align*}
c_i &\geq P_i A \xi^i\\
P\xi^i &\geq c + d\\
d_i &\geq P_i \omega^i\\
F\omega^i &\geq 0
\end{align*}
for $i = 1\dots r$. If Assumption~\ref{assump:P} holds, then
$H_iP=P_iA$ and $M_i F=P_i$ for some non-negative $H_i$ and
$M_i$. Substituting into the system $\vec{A}\vec{x} \geq 0$, 
\begin{align*}
c_i &\geq H_i P \xi^i\\
P \xi^i &\geq c + d\\
d_i &\geq M_i F \omega^i\\
F\omega^i &\geq 0,
\end{align*}
from which it follows that $d_i \geq 0$ and $c_i \geq H_i (c + d)$,
hence $c \geq H c$. But we also have that, if
Assumption~\eqref{assump:P} holds, then there exists some $q \in
\mbb{R}^r_{+}$ for which $0 \leq Hq \leq Hq + M g \leq q$. Applying
recursively, $0 \leq H^{n} q \leq Hq \leq q$, $H^n \geq 0$ because $H
\geq 0$, and therefore $\lim_{n \to \infty} H^n \geq 0$, if the limit
exists. In fact, because $HP = PA$, the nullspace of $P$ is
$A$-invariant and $P$ has rank $n$, then the eigenvalues of $H$ are
are subset of the eigenvalues of $A$; hence, $\lim_{n \to \infty} H^n
= 0$ because $\rho(A) < 1$. Then $c \geq H c \geq \lim_{n \to \infty}
H^n c = 0$. Consequently, $\vec{c}^\top \vec{x} = \sum_{i=1}^r c_i +
d_i \geq 0$. Therefore, $\mbb{D}$ is feasible if
Assumption~\ref{assump:P} holds.
\end{IEEEproof}

\section{Examples}
\label{sec:ex}

We consider the non-autonomous system
\begin{equation}
x^+ = 
\begin{bmatrix}
1 & 1 \\
0 & 1 
\end{bmatrix}
x+
\begin{bmatrix}
0.5 \\
1 
\end{bmatrix}
u + w,
\label{eq:exsys}
\end{equation}
with $w \in \mbb{W} = \{ w \in \mbb{R}^2: \| w\|_{\infty} \leq 0.1 \}$. This is
converted to the linear autonomous system~\eqref{eq:dyn} by use of a
state feedback control law $u = K x$.

\subsection{Computation of $(P,r)$-mRPI from selected inequalities}

First, we use the feedback matrix $K = [-0.4345, -1.0285]$,
corresponding to the infinite-horizon LQR solution with cost matrices
$Q = I$ and $R=1$. Note that in this example the mRPI set is not
finitely determined, and therefore an approximation is required.

\begin{figure}[t]
\centering\footnotesize
\subfloat[{$K=\left[-0.4345, -1.0285\right]$}]{\label{subfig:1}
%
%
%
%

\definecolor{mycolor1}{rgb}{0.831372559070587,0.815686285495758,0.7843137383461}
\definecolor{mycolor2}{rgb}{0.501960813999176,0.501960813999176,0.501960813999176}

\begin{tikzpicture}

\begin{axis}[%
width=0.5\textwidth,
height=0.4\textwidth,
xmin=-0.60115021972171, xmax=0.60115021972171,
xtick = {-0.5,-0.25,0,0.25,0.5},
xticklabels = {$-0.5$,$-0.25$,$0$,$0.25$,$0.5$},
xlabel={$x_1$},
ymin=-0.366081854763635, ymax=0.366081854763635,
ytick = {-0.2,0,0.2},
yticklabels = {$-0.2$,$0$,$0.2$},
ylabel={$x_2$},
area legend
]

\addplot [fill=lightgray,draw=none] table{
0.546500199747015 -0.16312457056567 0
0.448537071389799 -0.33280168614876 0
-0.260177041934735 -0.33280168614876 0
-0.546500199747015 0.16312457056567 0
-0.448537071389799 0.33280168614876 0
0.260177041934735 0.33280168614876 0
0.546500199747015 -0.16312457056567 0
};

\addlegendentry{$r=6$}

\addplot [fill=red,draw=none] table{
0.355701009681743 -0.036950553675871 0
0.36047858467399 -0.0516544175771283 0
0.36047858467399 -0.253494296339637 0
0.359105339641414 -0.254492017257175 0
0.357075619178756 -0.25515151341334 0
0.0767114768512201 -0.25515151341334 0
-0.0459415815171665 -0.215299118946247 0
-0.313850246362003 -0.0206520803153389 0
-0.355701009681743 0.0369505536758702 0
-0.36047858467399 0.051654417577128 0
-0.36047858467399 0.253494296339637 0
-0.359105339641414 0.254492017257175 0
-0.357075619178756 0.25515151341334 0
-0.0767114768512202 0.25515151341334 0
0.0459415815171664 0.215299118946247 0
0.313850246362002 0.0206520803153393 0
0.355701009681743 -0.036950553675871 0
};

\addlegendentry{$r=20$}



\addplot [fill=blue,draw=none] table{
0.33168593956748 -0.0163556692621447 0
0.348822253002328 -0.0386881477273979 0
0.351684903242533 -0.0436464033877316 0
0.35378210603691 -0.0487094988169638 0
0.35425011434258 -0.0504561295920885 0
0.35425011434258 -0.250456129592067 0
0.354126808385295 -0.251392731324234 0
0.354082320658849 -0.251558761779644 0
0.353482805630223 -0.252158276808271 0
0.353264438870177 -0.252325835516532 0
0.353097350315457 -0.252395045862014 0
0.352782320642256 -0.25247945780854 0
0.0988729593933565 -0.25247945780854 0
0.055082887916042 -0.246714385529021 0
0.0546725175634896 -0.246604427124457 0
0.0265123873194723 -0.234940119259192 0
-0.176752288143117 -0.117585204164124 0
-0.266976667050441 -0.0483536032548946 0
-0.331685939567424 0.0163556692620881 0
-0.348822253002367 0.0386881477274646 0
-0.351684903242533 0.0436464033877324 0
-0.35378210603691 0.0487094988169634 0
-0.35425011434258 0.0504561295920887 0
-0.35425011434258 0.250456129592066 0
-0.354126808385295 0.251392731324237 0
-0.35408232065885 0.251558761779644 0
-0.353482805630278 0.252158276808216 0
-0.353264438870141 0.252325835516546 0
-0.35309735031546 0.252395045862013 0
-0.352782320642255 0.25247945780854 0
-0.098872959393356 0.25247945780854 0
-0.0550828879160419 0.246714385529021 0
-0.0546725175634899 0.246604427124457 0
-0.0265123873194716 0.234940119259192 0
0.176752288143117 0.117585204164125 0
0.26697666705044 0.0483536032548958 0
0.33168593956748 -0.0163556692621447 0
};

\addlegendentry{$r=48$}

\addplot [fill=white,draw=none] table{
0.326629988753099 -0.0136169800002983 0
0.326725595899801 -0.0137190930196486 0
0.34075295864466 -0.0288956685359192 0
0.340810508254672 -0.0289811821678588 0
0.349059436934498 -0.0415276368246698 0
0.349084671091335 -0.0415662697858217 0
0.352692426738564 -0.0472289084647082 0
0.352695934474658 -0.0472514785805956 0
0.353058198451415 -0.0504783533299807 0
0.353131476218402 -0.0518846704116003 0
0.353131476218402 -0.251895416079711 0
0.351847532006817 -0.251960957707 0
0.151836786338732 -0.251960957707 0
0.151211002647818 -0.251952763541107 0
0.0540523760515463 -0.246259271065563 0
0.053015520001258 -0.245699553117051 0
-0.103544566933433 -0.158798235648827 0
-0.104030423905447 -0.15852657637547 0
-0.177316441953 -0.116474851755082 0
-0.177856157186297 -0.116040288068944 0
-0.258191081755068 -0.0504912808067601 0
-0.258439427251207 -0.0502879171284866 0
-0.295377330371579 -0.0196434118936422 0
-0.295588700279764 -0.0194212849292968 0
-0.326629988752573 0.0136169799997379 0
-0.326725595899877 0.0137190930197299 0
-0.340752958644656 0.0288956685359142 0
-0.340810508254702 0.0289811821679052 0
-0.349059436934636 0.0415276368248809 0
-0.349084671091335 0.0415662697858213 0
-0.352692426738563 0.047228908464707 0
-0.352695934474659 0.0472514785806096 0
-0.353058198451415 0.0504783533299816 0
-0.353131476218402 0.0518846704115994 0
-0.353131476218401 0.251895416079711 0
-0.351847532006821 0.251960957707 0
-0.151836786338732 0.251960957707 0
-0.151211002647842 0.251952763541107 0
-0.0540523760515425 0.246259271065562 0
-0.053015520000741 0.245699553116773 0
0.103544566933359 0.158798235648878 0
0.104030423904841 0.158526576375818 0
0.177316441952994 0.116474851755085 0
0.177856157186391 0.116040288068868 0
0.258191081755519 0.0504912808063915 0
0.258439427251179 0.0502879171285104 0
0.295377330371576 0.0196434118936454 0
0.295588700279703 0.0194212849293609 0
0.326629988753099 -0.0136169800002983 0
};

\addlegendentry{$\epsilon$-mRPI}

\end{axis}
\end{tikzpicture}%
}\\
\subfloat[{$K=\left[-0.0796, -0.4068\right]$}]{\label{subfig:2}\input{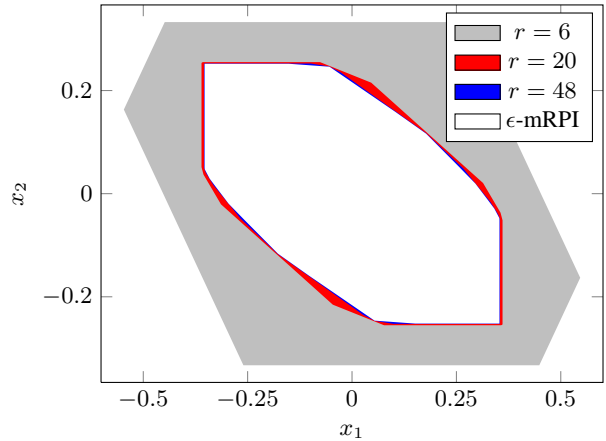}
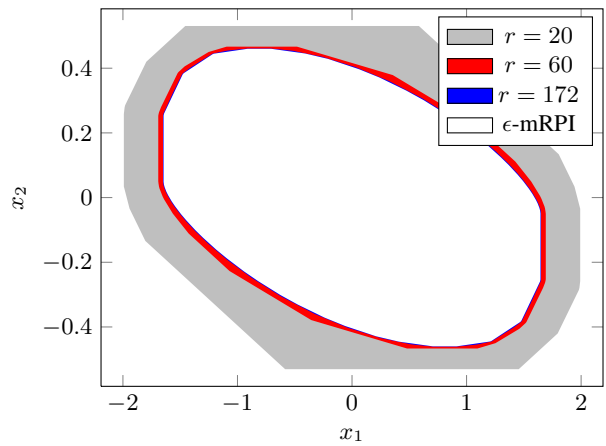}
\caption{Comparison of $(P,r)$-mRPI and $\epsilon$-mRPI sets for the
  system~\eqref{eq:exsys} with different feedback matrices $K$.}
\label{fig:1}
\end{figure}

Figure~\ref{fig:1}\subref{subfig:1} shows the $(P,r)$-mRPI sets generated from
$r=6$, $20$ and $48$ inequalities, wherein the $i$th row of $P$ is designed as
\begin{equation}
P_i = \begin{bmatrix}
\sin \left( \frac{2\pi (i-1)}{r} \right) & \cos \left( \frac{2\pi (i-1)}{r} \right)
\end{bmatrix},
\label{eq:Pdefin}
\end{equation}
\ie~so that $P x \leq 1$ is the $r$-sided regular polygon. Also shown
is the outer approximation to the mRPI, which is itself RPI,
computed using the algorithm of~\cite{RKK+05} and a tolerance
$\epsilon=10^{-4}$. This set, termed the $\epsilon$-mRPI set, is
defined by $48$ non-redundant inequalities.

Figure~\ref{fig:1}\subref{subfig:2} shows a similar comparison using $K
= [-0.0796,\allowbreak -0.4068]$, obtained as the LQR solution with $Q
= I$ and $R=100$. Now the $\epsilon$-mRPI ($\epsilon = 10^{-4}$)
comprises $172$ non-redundant inequalities, while the $(P,r)$-mRPI sets
computed using the proposed method are shown for $r=20$, $60$ and
$172$, again using~\eqref{eq:Pdefin} for $P$.

Table~\ref{tab:1} compares the computation times and number of operations for
computing the $(P,r)$-mRPI with those for obtaining the $\epsilon$-outer
approximation using the algorithm of~\cite{RKK+05}. For the latter,
the Multi-Parametric Toolbox v3.0~\cite{MPT3} was used for set
operations, with CPLEX 12.6 as the LP solver for support function
calculations. For the $(P,r)$-mRPI set computations (\ie~solving the LP),
CPLEX 12.6 was used as the LP solver. The platform was a $64$-bit
Intel Core i7-2600 at $3.40$~GHz with $8$~GB RAM. Times are reported
as the mean elapsed time over $100$ runs.

Comparison was also made with the iterative procedure of~\cite{RKC13}
for computing the $(P,r)$-mRPI. The iterative procedure is
\begin{equation*}
q_{k+1} = c(q_{k}) + d \ \text{with}\ q_0 = 0
\end{equation*}
for which $q_k \to q^*$ as $k \to \infty$. This was implemented in
\textsc{Matlab} using the MPT v3.0~\cite{MPT3} for support function
calculations (with CPLEX 12.6 as underlying LP solver). The function
$c(\cdot)$ was evaluated element by element at each iteration; that
is, as $r$ separate support function calculations. For the simplest
case considered of $K=\left[-0.4345, -1.0285\right]$ and $r=6$ (the
first row of Table~\ref{tab:1}), the number of iterations to
convergence (of $|q_{k+1} - q_k|$ to within a chosen tolerance of
$10^{-6}$) was $34$, which included the solving of $238$ LPs and took
a mean total time of $1.7$~seconds. At the other end of the scale, for
the most difficult problem considered ($K=\left[-0.0796,
  -0.4068\right]$ and $r=172$), the iterative procedure required $70$ iterations, the solving
of over $12000$ LPs, and took, on average, $90$~seconds. While these
times can, of course, be shortened by using optimized code, the
intention here is merely to report the times obtained using standard
computational tools.

\begin{table}[b]
\centering\footnotesize
\caption{Comparison of computation times and operations for $(P,r)$-mRPI
  and $\epsilon$-mRPI sets.}
\label{tab:1}
\begin{tabular}{lrrr}
\toprule
 & LPs solved & Minkowski sums & Mean time (s) \\
\midrule
\multicolumn{4}{c}{{$K=\left[-0.4345, -1.0285\right]$}}\\
$r=6$ & $1$ & $0$ & $0.005$ \\
$r=20$ & $1$ & $0$ & $0.007$ \\
$r=48$ & $1$ & $0$ & $0.019$ \\
$\epsilon$-mRPI~\cite{RKK+05} ($r=48$) & $369$ & $11$ & $2.9$ \\
\midrule
\multicolumn{4}{c}{{$K=\left[-0.0796, -0.4068\right]$}}\\
$r=20$ & $1$ & $0$ & $0.008$ \\
$r=60$ & $1$ & $0$ & $0.036$ \\
$r=172$ & $1$ & $0$ & $0.30$ \\
$\epsilon$-mRPI~\cite{RKK+05} ($r=172$) & $3250$ & $42$ & $25$ \\
\bottomrule
\end{tabular}
\end{table}

\subsection{Re-computing the $(P,r)$-mRPI set given $P$}

An interesting use of the method is when an RPI set for the system is
available, but is desired to be re-computed or modified; for example,
if the disturbance set changes. Potential applications of this include
``plug-and-play'' tube-based approaches to distributed MPC, wherein a
dynamic subsystems' disturbance set evolves over time as other
subsystems are added to and removed from the system of coupled
subsystems~\cite{RFF14}; in such situations, one needs a new RPI set
that takes into account the latest disturbance set. One could
re-compute from scratch a new RPI set, but it may be advantageous, in
the interests of computation time, to modify an existing RPI set
instead. In the context of the approach proposed here, the $P$ matrix of the known RPI set may be
used as a basis for computing the new RPI set.

For the system~\eqref{eq:exsys} with $K = [-0.4345, -1.0285]$ and
$\mbb{W} = \{ w \in \mbb{R}^2: |w|_\infty \leq 0.1 \}$, the $P$ matrix
is obtained as that of the $\epsilon$-mRPI set. For
$\epsilon=10^{-4}$, this comprises $48$ inequalities.  Now suppose the
disturbance set enlarges to
\begin{equation*}
\mbb{W} = \left\{ w \in \mbb{R}^2: \begin{bmatrix} -0.3\\-0.4 \end{bmatrix}  \leq w \leq \begin{bmatrix} 0.1\\0.2 \end{bmatrix} \right\}
\end{equation*}

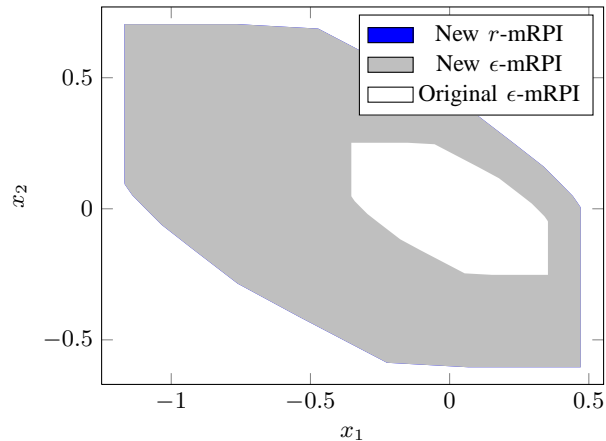
\begin{figure}[t]
\centering\footnotesize
%
%
%
%

\definecolor{mycolor1}{rgb}{0.501960784313725,0.501960784313725,0.501960784313725}

\begin{tikzpicture}

\begin{axis}[%
width=0.5\textwidth,
height=0.4\textwidth,
xmin=-1.25003027163392, xmax=0.55298992821618,
xlabel={$x_1$},
ymin=-0.670015569877017, ymax=0.770019790572959,
ylabel={$x_2$},
area legend
]

\addplot [fill=blue,draw=none] table{
0.442562062329898 0.0486954375290359 0
0.442678319545817 0.0485226895856469 0
0.45923549951776 0.0233395517715561 0
0.470073800468566 0.00632803874833114 0
0.470080886522386 0.00628244436734293 0
0.470808019716661 -0.000194511373745648 0
0.471028158376033 -0.00441932293296769 0
0.471028158376033 -0.604420256355952 0
0.468434434820218 -0.604552658419103 0
0.068434434820438 -0.604552658419103 0
0.0665544760217082 -0.604528041775219 0
-0.224906197519041 -0.587448455434247 0
-0.227000772921333 -0.586317757063101 0
-0.540104124266233 -0.412524459752668 0
-0.541563719811348 -0.411708349893139 0
-0.761410303998606 -0.285559757524855 0
-0.762500594450615 -0.284681885994546 0
-0.923161811527356 -0.153590914770698 0
-0.923907882904689 -0.152979976292215 0
-1.03471581113304 -0.0610512567438854 0
-1.03514280406869 -0.0606025332457934 0
-1.09722204560175 0.00547044662291569 0
-1.09750926545094 0.00577721120097642 0
-1.13958915826945 0.0513045624709717 0
-1.13970541548521 0.0514773104141124 0
-1.15620238648996 0.0765688715997647 0
-1.15627819411406 0.0766849314711409 0
-1.16710089640814 0.0936719612517174 0
-1.16710798246187 0.0937175556321536 0
-1.16783511565621 0.100194511373766 0
-1.1680552543156 0.104419322933401 0
-1.1680552543156 0.704420256355947 0
-1.16546153075969 0.704552658419103 0
-0.765461530760142 0.704552658419103 0
-0.763581571961198 0.704528041775215 0
-0.472120898420659 0.687448455434256 0
-0.470026323013791 0.68631775706064 0
-0.156922971671111 0.51252445975144 0
-0.155463376128022 0.511708349893043 0
0.0643832080589292 0.385559757524935 0
0.0654734985100878 0.384681885995311 0
0.226134715591573 0.253590914767618 0
0.226880786965421 0.252979976291989 0
0.337688715193655 0.1610512567437 0
0.338115708131092 0.160602533243726 0
0.400194949659178 0.0945295533803093 0
0.400482169511518 0.0942227887988809 0
0.442562062329898 0.0486954375290359 0
};

\addlegendentry{New $r$-mRPI}

\addplot [fill=lightgray,draw=none] table{
0.442570960495283 0.0486941172251684 0
0.442686060529618 0.0485230887512953 0
0.459184034608744 0.0234300019096988 0
0.459259737614693 0.0233141022064985 0
0.470083081129144 0.00632606598323407 0
0.470090096650965 0.00628092543210327 0
0.470814629730389 -0.00017286972582685 0
0.47103446458663 -0.00439185081900811 0
0.47103446458663 -0.604428332947325 0
0.468466557996125 -0.604559417129291 0
0.068442236577168 -0.604559417129291 0
0.0665648722226883 -0.604534834457699 0
-0.224913069707456 -0.58745423618965 0
-0.226986796480282 -0.586334792372221 0
-0.540109185619581 -0.412530927812833 0
-0.541566766845684 -0.411715944228053 0
-0.761426376447807 -0.285559877839846 0
-0.762505814551267 -0.284690744318579 0
-0.923176800400718 -0.15359180229801 0
-0.923921842159309 -0.152981706947581 0
-1.03473633550806 -0.0610475408294838 0
-1.03515907831498 -0.0606032837580553 0
-1.09724209448494 0.00547371358094335 0
-1.09752891795497 0.00578005480705032 0
-1.13961130391302 0.0513101034707691 0
-1.13972640394731 0.0514811319445795 0
-1.15622437802688 0.0765742187868462 0
-1.15630008103235 0.0766901184893258 0
-1.16712342454688 0.0936781547127078 0
-1.1671304400687 0.0937232952638299 0
-1.16785497314812 0.10017709042176 0
-1.16807480800437 0.104396071514962 0
-1.16807480800437 0.704432553643265 0
-1.16550690141385 0.704563637825232 0
-0.765482579994907 0.704563637825232 0
-0.76360521564044 0.704539055153641 0
-0.472127273710279 0.68745845688559 0
-0.470053546937538 0.686339013068207 0
-0.15693115779832 0.512535148508865 0
-0.155473576572014 0.511720164923971 0
0.0643860330301552 0.385564098535738 0
0.0654654711325399 0.384694965015337 0
0.226136456983213 0.253596022993761 0
0.226881498741452 0.25298592764362 0
0.337695992090728 0.161051761524999 0
0.338118734897423 0.160607504453801 0
0.400201751067034 0.0945305071151807 0
0.400488574537336 0.0942241658887851 0
0.442570960495283 0.0486941172251684 0
};

\addlegendentry{New $\epsilon$-mRPI}

\addplot [fill=white,draw=none] table{
0.326629988753109 -0.0136169800003086 0
0.326725595899797 -0.0137190930196437 0
0.34075295864466 -0.0288956685359184 0
0.340810508254707 -0.0289811821679105 0
0.349059436934585 -0.0415276368248012 0
0.349084671091299 -0.0415662697857649 0
0.352692426738563 -0.0472289084647078 0
0.352695934474658 -0.0472514785805959 0
0.353058198451415 -0.0504783533299803 0
0.353131476218402 -0.0518846704116009 0
0.353131476218402 -0.251895416079711 0
0.351847532006818 -0.251960957707 0
0.151836786338729 -0.251960957706999 0
0.151211002647818 -0.251952763541107 0
0.0540523760515452 -0.246259271065563 0
0.0530155200013094 -0.24569955311708 0
-0.103544566933433 -0.158798235648827 0
-0.104030423905439 -0.158526576375474 0
-0.177316441953001 -0.116474851755081 0
-0.177856157186292 -0.116040288068949 0
-0.258191081755165 -0.0504912808066811 0
-0.258439427251209 -0.0502879171284848 0
-0.295377330371581 -0.019643411893641 0
-0.295588700279751 -0.0194212849293104 0
-0.326629988752585 0.0136169799997501 0
-0.326725595899872 0.0137190930197253 0
-0.34075295864466 0.0288956685359185 0
-0.340810508254651 0.0289811821678281 0
-0.34905943693465 0.0415276368249035 0
-0.349084671091338 0.0415662697858269 0
-0.352692426738563 0.0472289084647068 0
-0.35269593447466 0.0472514785806115 0
-0.353058198451415 0.0504783533299785 0
-0.353131476218402 0.0518846704116008 0
-0.353131476218402 0.251895416079711 0
-0.351847532006823 0.251960957706999 0
-0.15183678633873 0.251960957706999 0
-0.15121100264784 0.251952763541107 0
-0.0540523760515425 0.246259271065562 0
-0.0530155200007413 0.245699553116773 0
0.103544566933359 0.158798235648878 0
0.104030423904835 0.158526576375821 0
0.177316441952994 0.116474851755085 0
0.177856157186388 0.11604028806887 0
0.258191081755505 0.050491280806403 0
0.258439427251187 0.0502879171285038 0
0.29537733037158 0.0196434118936424 0
0.29558870027967 0.0194212849293966 0
0.326629988753109 -0.0136169800003086 0
};

\addlegendentry{Original $\epsilon$-mRPI}

\end{axis}
\end{tikzpicture}%
\caption{Comparison of $(P,r)$-mRPI and $\epsilon$-mRPI sets for the
  system~\eqref{eq:exsys} with $K=[-0.4345, -1.0285]$ and different disturbance sets.}
\label{fig:2}
\end{figure}
Figure~\ref{fig:2} shows the $(P,r)$-mRPI and $\epsilon$-mRPI sets
based on the new disturbance set, using for the former the $P$ matrix
from the old $\epsilon$-mRPI set. The $(P,r)$-mRPI set, computed in
$0.03$~s using the proposed method, is visually indistinguishable from
the new $\epsilon$-mRPI set.

\section{Conclusions}
\label{sec:conc}
A procedure for computing a polytopic robust positively invariant set
for a linear uncertain system has been presented. The method, which
requires the solution of a single LP, obtains the an RPI set that is
the smallest among those represented by a finite number inequalities
with pre-defined normal vectors, and offers an alternative method of
computation to the iterative procedure of~\cite{RKC13}. Existence and
uniqueness of a solution has been established. The practicality of the
approach has been demonstrated via examples.

\section*{Acknowledgments}

The author would like to thank the anonymous reviewers for their
constructive and useful comments.

\bibliography{extracted}

\begin{thebibliography}{10}
\providecommand{\url}[1]{#1}
\csname url@samestyle\endcsname
\providecommand{\newblock}{\relax}
\providecommand{\bibinfo}[2]{#2}
\providecommand{\BIBentrySTDinterwordspacing}{\spaceskip=0pt\relax}
\providecommand{\BIBentryALTinterwordstretchfactor}{4}
\providecommand{\BIBentryALTinterwordspacing}{\spaceskip=\fontdimen2\font plus
\BIBentryALTinterwordstretchfactor\fontdimen3\font minus
  \fontdimen4\font\relax}
\providecommand{\BIBforeignlanguage}[2]{{%
\expandafter\ifx\csname l@#1\endcsname\relax
\typeout{** WARNING: IEEEtran.bst: No hyphenation pattern has been}%
\typeout{** loaded for the language `#1'. Using the pattern for}%
\typeout{** the default language instead.}%
\else
\language=\csname l@#1\endcsname
\fi
#2}}
\providecommand{\BIBdecl}{\relax}
\BIBdecl

\bibitem{KG98}
I.~Kolmanovsky and E.~G. Gilbert, ``Theory and computation of disturbance
  invariant sets for discrete-time linear systems,'' \emph{Mathematical
  Problems in Engineering}, vol.~4, pp. 317--367, 1998.

\bibitem{Blanchini99}
F.~Blanchini, ``Set invariance in control,'' \emph{Automatica}, vol.~35,
  no.~11, pp. 1747--1767, 1999.

\bibitem{RKK+05}
S.~V. Rakovi\'c, E.~C. Kerrigan, K.~I. Kouramas, and D.~Q. Mayne, ``Invariant
  approximations of the minimal robust positively invariant set,'' \emph{IEEE
  Transactions on Automatic Control}, vol.~50, no.~4, pp. 406--410, April 2005.

\bibitem{RKM+07}
S.~V. Rakovi\'c, E.~C. Kerrigan, D.~Q. Mayne, and K.~I. Kouramas, ``Optimized
  robust control invariance for linear discrete-time systems: Theoretical
  foundations,'' \emph{Automatica}, vol.~43, no.~5, pp. 831--841, May 2007.

\bibitem{MSR05}
D.~Q. Mayne, M.~M. Seron, and S.~V. Rakovi\'c, ``Robust model predictive
  control of constrained linear systems with bounded disturbances,''
  \emph{Automatica}, vol.~41, no.~2, pp. 219--224, 2005.

\bibitem{RKC13}
S.~V. Rakovi\'c, B.~Kouvaritakis, and M.~Cannon, ``Equi-normalization and exact
  scaling dynamics in homethetic tube model predictive control,'' \emph{Systems
  \& Control Letters}, vol.~62, no.~2, pp. 209--217, February 2013.

\bibitem{RFF14}
S.~Riverso, M.~Farina, and G.~Ferrari-Trecate, ``Plug-and-play model predictive
  control based on robust control invariant sets,'' \emph{Automatica}, vol.~50,
  pp. 2179--2186, 2014.

\bibitem{HC01}
J.-C. Hennet and E.~B. Castelan, ``Invariant polyhedra and control of
  large-scale systems,'' in \emph{Proceedings of the 9th IFAC Symposium on
  Large Scale Systems: Theory and Applications}, 2001.

\bibitem{Schneider93}
R.~Schneider, \emph{Convex bodies: the {B}runn-{M}inkowski theory}, ser.
  Encyclopedia of Mathematics and its Applications, G.-C. Rota, Ed.\hskip 1em
  plus 0.5em minus 0.4em\relax Cambridge University Press, 1993, vol.~44.

\bibitem{MPT3}
M.~Herceg, M.~Kvasnica, C.~N. Jones, and M.~Morari, ``{Multi-Parametric Toolbox
  3.0},'' in \emph{Proc.~of the European Control Conference}, Z\"urich,
  Switzerland, July 17--19 2013, pp. 502--510,
  \url{http://control.ee.ethz.ch/~mpt}.

\end{thebibliography}

\end{document}